# Epitaxial growth and magnetic properties of NiMnAs film on GaAs substrate

J. L. Ma[1,2], H. L. Wang[1,2*], X. M. Zhang[3], S. Yan[3], W. S. Yan[4], and J. H. Zhao[1,2,5]

[1]*State Key Laboratory of Superlattices and Microstructures, Institute of Semiconductors, Chinese Academy of Sciences, P. O. Box 912, Beijing 100083, China*

[2]*College of Materials Science and Opto-Electronic Technology, University of Chinese Academy of Sciences, Beijing 100049, China*

[3]*Shanghai Synchrotron Radiation Facility, Institute of Applied Physics, Chinese Academy of Sciences, Shanghai 201204, China*

[4]*National Synchrotron Radiation Laboratory, University of Science and Technology of China, Hefei 230029, China*

[5]*CAS Center for Excellence in Topological Quantum Computation, University of Chinese Academy of Sciences, Beijing 100190, China*

Single-phase $Ni_{0.46}Mn_{0.54}As$ films with strained $C1_b$ symmetry have been successfully grown on GaAs (001) substrates by molecular-beam epitaxy. The epitaxial relationship between the film and the substrate has been studied using synchrotron radiation, and a preferred configuration of (110)-orientated $Ni_{0.46}Mn_{0.54}As$ on (001)-orientated GaAs was revealed. In addition, the magnetic properties of the films were found to be significantly influenced by the growth temperature. The optimized growth temperature is determined to be ~370 °C, for which relatively high Curie temperature, large saturation magnetization and coercive field, as well as the pronounced in-plane magnetic anisotropy were obtained. According to the results of X-ray absorption spectroscopy, the above phenomenon can be attributed to the variation of the local electronic structure of the Mn atoms. Our work provides useful information for the further investigations of NiMnAs, which is theoretically predicted to host robust half-metallicity.

Corresponding authors: *allen@semi.ac.cn

Heusler alloys, due to their rich physical connotation and potential applications, have been extensively investigated [1,2]. Some intriguing physical phenomena, such as helimagnetism, non-collinear magnetism, and/or the superconducting behavior were successfully identified in this kind of alloys [1,3-8]. In addition, the shape memory effect accompanied by the thermoelastic martensitic transformation in some Heusler alloys also renders them the ideal functional materials for practical applications [9-11]. Among the numerous Heusler alloys, we notice that a series of materials with half-metallicity have received special attention, satisfying the requirements of high spin polarization and high Curie temperature ($T_C$) in some spin-based devices [12]. For instance, as the widely-studied half-metallic Heusler alloy, NiMnSb, has been extensively employed in the studies of tunnel magnetoresistive junction [13,14], spin-injection devices [15], giant magnetoresistance devices [16], and spin-valve structures [16,17]. Therefore, as an important branch of the researches that promotes the development of spintronics, seeking for new half-metallic Heusler alloy is always in the ascendant.

Specifically, half-Heusler alloy NiMnAs with $C1_b$ structure was theoretically predicted to be half-metallic, with its $T_C$ much higher than room temperature (~840 K) [18,19]. More importantly, as a compound isoelectronic with NiMnSb, NiMnAs has some unique features that are absent in NiMnSb: while the half-metallicity of NiMnSb was found to be sensitive to the interface of heterojunction and the expansion and/or compression of its crystal lattice [18-22], the influences of the interface and lattice distortion on NiMnAs were predicted to be relatively weaker [18,23]. Furthermore, the minority spin gap of NiMnAs is larger (~ 0.69 eV) than that of NiMnSb [24], guaranteeing a much more robust half-metallicity against external perturbation. However, to our best knowledge, cubic NiMnAs has not been synthesized before, due to its tendency of crystallizing in the hexagonal or orthorhombic structures [3,25,26]. For the above reasons, the experimental investigations on NiMnAs with $C1_b$ structure are immensely limited, which makes the synthesis and fundamental characterizations of this material a meaningful and challenging work.

In this letter, we succeed in growing high-quality $Ni_{0.46}Mn_{0.54}As$ films with strained

$C1_b$ crystal structure using molecular-beam epitaxy. The $Ni_{0.46}Mn_{0.54}As$ (110) plane was found to be parallel with the GaAs (001) plane. In addition, the growth temperature ($T_s$) dependence of the magnetic properties for the $Ni_{0.46}Mn_{0.54}As$ films has been systematically studied, in which an optimized growth temperature was determined to be ~370 °C. For the sample grown at this optimized $T_s$, relatively high $T_C$, large saturation magnetization ($M_s$) and coercive field ($H_c$), as well as pronounced in-plane magnetic anisotropy (MA) were obtained. Based on this newly-synthesized film, rich spin-dependent physics in NiMnAs can be explored, which will facilitate further investigations of this material and its relevant devices.

A series of NiMnAs films were epitaxied on GaAs (001) substrates at different temperatures, and the layer structure is schematically shown in Fig. 1(a). In detail, a 150 nm thick undoped GaAs buffer layer was firstly grown on GaAs (001) substrate after the desorption of the natural oxidation layer at ~580 °C. Subsequently, the substrate was cooled down to room temperature and a ~0.6 nm thick low-temperature NiMnAs layer was deposited to eliminate the influences of interfacial reaction. Meanwhile, the sample surface was monitored using *in situ* reflection high-energy electron diffraction (RHEED), which presented the halo pattern after the deposition of the low-temperature layer. We annealed this layer until the RHEED patterns changed from halo to streaky, indicating the transition of the film from amorphous to crystalline. Afterwards, the NiMnAs films with a fixed thickness of ~30 nm were grown at various temperatures (250, 300, 330, 370, and 400 °C). At this growth stage, RHEED always showed streaky (1×2) patterns [Fig. 1(b)], suggesting a layer-by-layer growth mode and the high quality of the films. In order to prevent the NiMnAs film from being oxidized, a 3 nm thick GaAs capping layer was finally grown. The thickness and growth rate of the NiMnAs films were calibrated by X-ray reflectivity (XRR) measurement [Fig. 1(c)], which also implied a relatively smooth interface between the NiMnAs film and the GaAs substrate, consistent with the observed RHEED patterns [Fig. 1(b)]. In addition, energy dispersive spectrometer (EDS) was used to check the chemical composition of the films, and the result of Ni:Mn~0.46:0.54 is shown in Fig. 1(d), a good representation of the stoichiometric NiMnAs films.

We firstly used a laboratory X-ray diffractometer ($\lambda_1 \sim 1.5406$ Å) to unveil the crystal structure of the films. Figure 2(a) illustrates the room-temperature $\omega$-$2\theta$ diffraction results from five $Ni_{0.46}Mn_{0.54}As$ films with different growth temperatures. For each diffraction curve, three peaks can be seen, among which the peaks at $2\theta \sim 31.6°$ and $2\theta \sim 66.1°$ correspond to the reflections of GaAs (002) and (004) crystallographic planes, respectively. The relatively weak peak between the above two peaks should be attributed to the X-ray diffraction of the $Ni_{0.46}Mn_{0.54}As$ film, which slightly shifts with the variation of the growth temperature, implying the change of the lattice relaxation. Except for these three peaks, no other peaks can be found in the broad $\theta$ range, which suggests that all $Ni_{0.46}Mn_{0.54}As$ films are high-quality single-crystalline films without observable second phase. Considering that the lattice constant of NiMnAs with $C1_b$ structure is similar with GaAs [19,23,27], and the reflection position of the film ($2\theta \sim 44°$) is very close to that of GaAs (220) plane, it is hence reasonable to ascribe this peak to the reflection of quasi-cubic $Ni_{0.46}Mn_{0.54}As$ (220) crystallographic plane. In other words, the configuration of (110)-orientated $Ni_{0.46}Mn_{0.54}As$ on (001)-orientated GaAs probably establishes between the film and the substrate.

To confirm the above special epitaxial relationship, an experimental setup based on synchrotron radiation was utilized, in which the diffraction data was collected with the sample rotating under a grazing incident high-energy X-ray ($\lambda_2 \sim 0.6895$ Å). Figure 2(b) presents the typical result of the $Ni_{0.46}Mn_{0.54}As$ film grown at $T_s \sim 370$ °C, in which the sample was slowly rotated around the surface normal (characterized by angle $\varphi$) from 0° to 180°. We define $\varphi=0°$ as one of the angles corresponding to the diffraction peaks of the film {004} planes. It can be seen that one GaAs {004} diffraction peak appears at $\varphi=45°$, which, together with the above XRD data, unambiguously verify the (quasi-)cubic structure of $Ni_{0.46}Mn_{0.54}As$ and reveal the epitaxial configuration of (110)-orientated $Ni_{0.46}Mn_{0.54}As$ film on (001)-orientated GaAs substrate [see the inset in Fig. 2(b)]. Similar epitaxial relationship was also found between $Ni_2MnIn$ Heusler film and InAs substrate regardless of the large lattice mismatch between them, which was ascribed to the interaction at the heterostructure interface [28,29].

The Quantum Design superconducting quantum interference device (SQUID)

magnetometer was utilized to characterize the basic magnetic properties of the Ni$_{0.46}$Mn$_{0.54}$As films. Figure 3(a) shows the temperature dependence of the magnetization for the Ni$_{0.46}$Mn$_{0.54}$As films grown at different temperatures, in which the magnetization component projected to GaAs [110] direction was measured. It can be seen that $T_C$ is significantly affected by the growth temperature and varies from 45 K to 295 K, among which the highest $T_C$ is obtained at $T_s$~370 °C. Note that, all of the $T_C$ values are much smaller than that predicted by the theoretical work (~840 K) [18,19]. In addition, we also find that the curves with the growth temperature higher than 250 °C all exhibit a hump within a large temperature range, which might be induced by the spin reorientation transition (SRT) effect. Similar phenomenon was observed in (Mn$_{1-x}$Ni$_x$)$_{65}$Ga$_{35}$, in which the SRT effect was suggested to be resulting from the presence of a temperature-sensitive helical spin structure [30]. Figure 3(b) presents the magnetic hysteresis loops (-3 T~3 T) for the above five samples at 5 K, in which the external magnetic field was applied along the GaAs [110] direction. It can be seen that the in-plane ferromagnetic behavior featuring a relatively rectangular loop appears in the samples with $T_s$ higher than 300 °C, and the most pronounced in-plane magnetic anisotropy arises at $T_s$~370 °C. In addition, the values of $H_c$ and $M_s$ are extrapolated from these loops, which are summarized in Fig. 3(c). Evidently, the largest $H_c$ and $M_s$ are also obtained at $T_s$~370 °C, as indicated by the dashed ellipse. Therefore, the optimized growth temperature associated with the largest values of $T_C$, $H_c$, $M_s$ and the most pronounced MA is determined to be ~370 °C. To further clarify the magnetic anisotropy, the magnetic hysteresis loops with the external magnetic field along the GaAs [-110], [110] and [001] directions were measured at 5 K, in which the Ni$_{0.46}$Mn$_{0.54}$As sample grown at $T_s$~370 °C was selected [see Fig. 3(d)]. As we can see, compared with the out-of-plane direction, a fourfold in-plane magnetic anisotropy is dominant in our sample. Deeper insights of the $T_s$-dependent magnetic properties are critical for effectively tailoring the magnetism of Ni$_{0.46}$Mn$_{0.54}$As, which motivate us to unravel the relevant mechanism.

Generally, macroscopic magnetic properties are closely associated with the microscopic electronic structure, which can be detected by the X-ray absorption spectroscopy (XAS). Taking the advantages of the intense and tunable X-ray beams of synchrotron radiation, XAS characteristic spectra of Mn and Ni elements were obtained. Figure 4(a) shows the room-temperature Mn 2$p$ XAS spectra of the Ni$_{0.46}$Mn$_{0.54}$As films with various growth temperatures, in which the energy of the X-ray beams was continuously changed. In this energy range, there are two types of XAS line shape, i.e. the shoulder structure (for the samples grown at $T_s$~250 °C) and the multiplet structure (for the samples grown at $T_s$~300, 330, 370, and 400 °C), respectively. This drastic difference for the XAS results suggests the evolution of the local electronic structure of the Mn atoms, which may further affect the magnetic properties of the Ni$_{0.46}$Mn$_{0.54}$As films. In detail, the shoulder structure without apparent multiplets reflects the initial metallic Mn $L$-edge shape, and can be identified in the Mn-based metallic systems, such as MnSb and pure Mn metal [31-34]. With the increase of the itinerancy of the Mn 3$d$ electrons, the spectrum will usually be broadened [32,33]. As for the multiplet structure, it probably indicates the covalent bonding of the Mn and As atoms, according to the analyses of the XAS spectra in NiMnSb [31]. Additionally, this structure is predicted to be associated with the Mn 3$d^5$ ground state configuration, as suggested by the theoretical calculation for Mn 2$p$ XAS spectra [32-34]. To determine the role of the growth temperature in the local structure of the Ni atoms, the measurement of Ni 2$p$ XAS spectra for the above set of Ni$_{0.46}$Mn$_{0.54}$As films was also carried out, as shown in Fig. 4(b). It is evident that the corresponding XAS line shapes are almost unchanged, indicating the variation of the growth temperature has no observable influence on the local electronic structure of the Ni atoms, which is in sharply contrary to the Mn 2$p$ XAS spectra. Correspondingly, it can be deduced that the magnetism in the Ni$_{0.46}$Mn$_{0.54}$As films is mainly contributed by the Mn atoms, which is analogous to NiMnSb [33,35]. Therefore, we conclude that the growth temperature dependence of the magnetic properties in Ni$_{0.46}$Mn$_{0.54}$As is related to the local electronic structure of the Mn atoms, while it is independent on that of the Ni atoms.

In summary, we have successfully synthesized the single-phase Ni$_{0.46}$Mn$_{0.54}$As films with strained $C1_b$ symmetry. A preferred epitaxial relationship of Ni$_{0.46}$Mn$_{0.54}$As (110)/GaAs (001) has been revealed, regardless of the large lattice mismatch for this configuration. In addition, the growth temperature is found to have remarkable effect on the magnetic properties of the $Ni_{0.46}Mn_{0.54}As$ films, which can be attributed to the variation of the local electronic structure of the Mn atoms, as verified by the XAS measurements. Our work is fundamental for the further studies of NiMnAs, and a perspective of the high-performance spin-based devices based on this material is expected.

This work was supported by MOST (Grant No. 2017YFB0405701), and NSFC (Grants Nos. U1632264 and 11706374), and the Key Research Project of Frontier Science of Chinese Academy of Science (Grant No. QYZDY-SSW-JSC015).

# Figures and captions

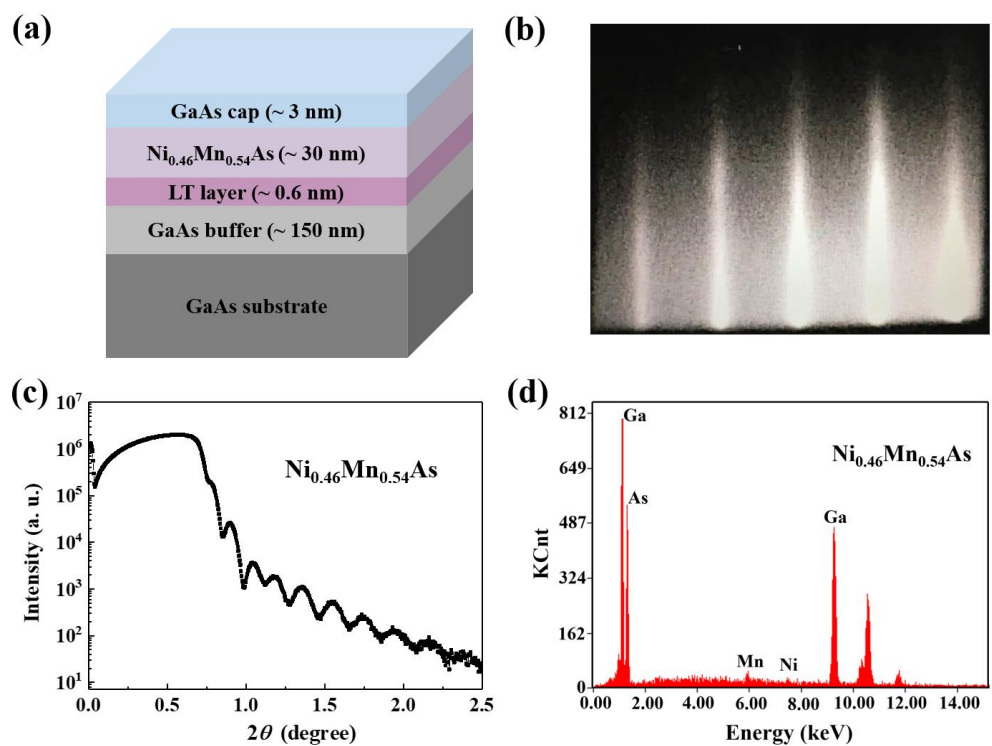

**FIG. 1** (a) Schematic diagram of layer structure of epitaxial Ni$_{0.46}$Mn$_{0.54}$As films. (b) Typical RHEED pattern of Ni$_{0.46}$Mn$_{0.54}$As film grown on GaAs (001) substrate. (c) XRR data of the Ni$_{0.46}$Mn$_{0.54}$As film with the thickness of ~30 nm. (d) EDS spectrum of the Ni$_{0.46}$Mn$_{0.54}$As film grown at 370 °C.

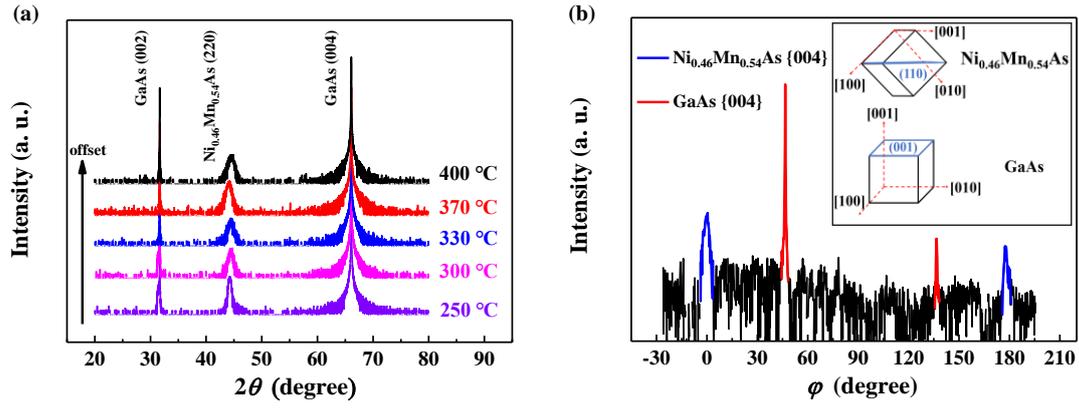

**FIG. 2** (a) room-temperature XRD ($\lambda_1 \sim 1.5406$ Å) data of the Ni$_{0.46}$Mn$_{0.54}$As films with various growth temperatures. (b) In-plane X-ray diffraction result of Ni$_{0.46}$Mn$_{0.54}$As film using the synchrotron radiation ($\lambda_2 \sim 0.6895$ Å). The inset schematically shows the configuration of (110)-orientated Ni$_{0.46}$Mn$_{0.54}$As film on (001)-orientated GaAs substrate.

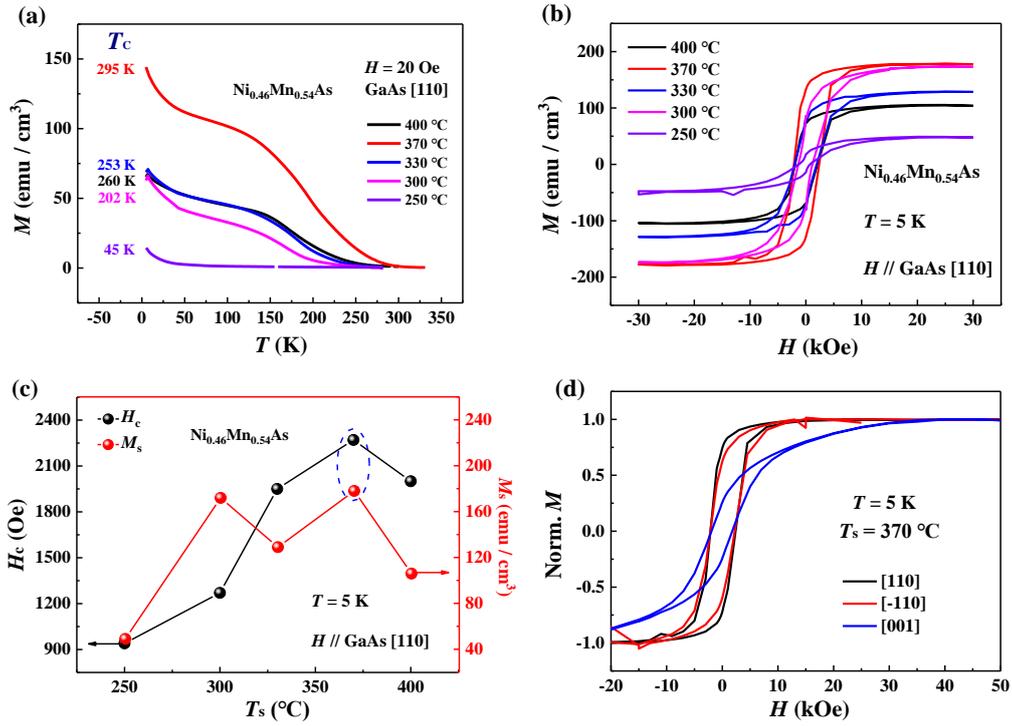

**FIG. 3** (a) Temperature dependence of the magnetization for the $Ni_{0.46}Mn_{0.54}As$ films grown at different temperatures, in which the magnetization component was measured along the GaAs [110] direction. (b) Magnetic hysteresis loops for the $Ni_{0.46}Mn_{0.54}As$ films deposited at various temperatures. These hysteresis loops were measured along the GaAs [110] direction at 5 K. (c) Growth temperature dependence of the coercive field and saturation magnetization extrapolated from fig. 3(b). (d) Normalized magnetic hysteresis loops for the $Ni_{0.46}Mn_{0.54}As$ film with $T_s$~370 °C at 5 K, in which the applied magnetic field is along the GaAs [110] (black), [-110] (red) and [001] (blue) directions, respectively.

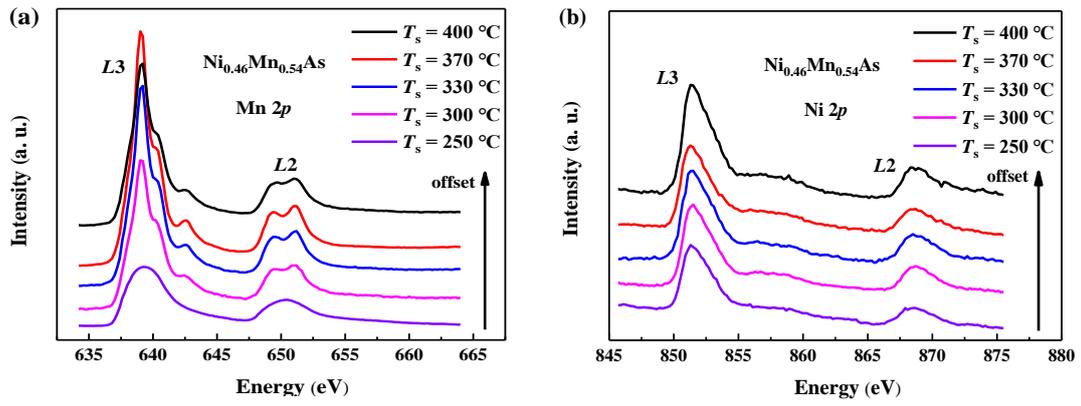

**FIG. 4** Room-temperature (a) Mn 2$p$, and (b) Ni 2$p$ XAS spectra of the Ni$_{0.46}$Mn$_{0.54}$As films with various growth temperatures.